\documentclass[twocolumn,showpacs,aps,eqsecnum]{revtex4}
\usepackage{graphicx,epsfig}
\usepackage[english]{babel}
\usepackage{subfigure}
\usepackage{graphicx}
\usepackage{latexsym}
\usepackage[ps2pdf]{hyperref}
\begin{document}
\baselineskip 8truemm
\title{Properties of a protoneutron star in the Effective Field Theory}
\author{Ilona Bednarek and Ryszard Manka} \email{manka@us.edu.pl}
\affiliation{Department of Astrophysics and Cosmology., Institute of Physics,
University of Silesia, Uniwersytecka 4, 40-007 Katowice, Poland}
\date{\today}
\begin{abstract}
The complete form of the equation of state of strangeness rich
proto-neutron and neutron star  matter has been obtained. The
currently obtained lower value of the $U_{\Lambda}^{(\Lambda )}$
potential at the level of 5 MeV permits the existence of
additional parameter set which reproduces the weaker $\Lambda
\Lambda$ interaction. The effects of the strength of
hyperon-hyperon interactions on the equations of state constructed
for the  chosen parameter set have been analyzed.  It has been
shown that replacing the strong $Y-Y$ interaction model by the
weak one introduces large differences in the composition of a
proto-neutron star matter both in the strange and non-strange
sectors. Also concentrations of neutrinos have been significantly
altered in proto-neutron star interiors. The performed
calculations have indicated that the change of the hyperon-hyperon
coupling constants affects the value of the proto-neutron star
maximum mass.
\end{abstract}
%\date{\today}
\maketitle
\section{Introduction}
The mass of a star is a key factor in determining  the stellar
evolution. Low-mass stars  developing a degenerate core can evolve
to the white dwarf stage. Evolutionary models of massive stars,
with the main sequence mass exceeding $8 M_{\odot}$ show that
their evolution proceeds through  sequences of consecutive stages
of nuclear burning until the iron core is formed. The formation of
${}^{56}Fe$ nucleus, with the maximum binding energy per nucleus,
indicates the beginning of the end of massive star's life as a
normal star. The interior composition of a massive star in a very
advanced stage of evolution prior to core collapse reveal a shell
structure. Each shell has different chemical composition of
gradually heavier elements. After the nuclear fuel is exhausted in
a star, nuclear fusion in the central part of the star can no
longer supply enough energy to sustain a high thermal pressure.
The obtained theoretical models of the evolved massive stars
indicate that they develop central iron cores of mass $\sim 1.5
M_{\odot}$ with the electron gas as the dominant source of
pressure. Both thermal pressure and the pressure of degenerate
electron gas attempt to support the core and determine  the following equation of
state of the innermost region of the star  \cite{Shapiro}
\begin{equation}
\frac{P}{\rho}\simeq
\frac{Y_ek_BT}{m_B}+K_{\Gamma}Y_e^{\Gamma}\rho^{\Gamma -1}.
\end{equation}
In this equation $Y_e=n_e/n_b$ is the electron concentration,
$\rho$ is the density, $k_B$ is Boltzmann's constant and $\Gamma$ is
the adiabatic exponent.
 Owing to the
high temperature of the core the adiabatic exponent is close to
the critical value 4/3 and the core become dynamically unstable.
\newline
The collapse of the core is accelerated by the combination of two
mechanisms \cite{Suzuki}: reactions of electron capture and
photodisintegration of iron-peak nuclei into $\alpha$ particles
${}^{56}Fe\rightarrow 13\alpha + 4n$. Iron dissociation is an
endothermic reaction, the energy cost equals 124.4 MeV, which comes
from the kinetic energy of nuclei and electrons and decreases
their thermal pressure. The resulting pressure deficit is
compensated by a further contraction of the core. Electron capture
occurs on nuclei and on free protons, diminishes the electron
fraction $Y_e$  and also leads to pressure deficit. As a
consequence the core collapse sets in and the star undergoes
supernova explosion,  ejecting a large part of its mass. Dynamical
conditions in the central region of the star split the core into
two separate parts: homologously collapsing inner core and the
free falling outer one.
 After the inner core reaches the density comparable with the
nuclear density, the equation of state stiffens, nuclear forces are
expected to resist compression and stop the collapse. The bounced
inner core settles into a hydrostatic configuration in the
dynamical time scale of milliseconds. At the boundary between the
inner core and the supersonically falling outer core  a pressure
wave, which originates from the center, become a shock wave.
 The shock wave travels into the outer core through
the material that is falling towards the center. This outgoing
shock dissociates the iron nuclei into protons and neutrons.
 The shocked region attains
high temperature  at comparatively  low densities. Under these
conditions electron degeneracy is not high and relativistic
positrons are thermally created. The thermal energy is carried out
by neutrinos which are produced in the following reactions
$e^++e^-\rightarrow \nu +\bar{\nu}$,
$e^++n\rightarrow\bar{\nu}+p$. Within the shock and in the region
behind the shock the velocity of the infalling matter is
significantly reduced and the falling matter eventually settles
onto the newly formed protoneutron star. Electron neutrinos
produced in $e^-+p\rightarrow \nu_e+n$ carry out electron-type
lepton number. Deleptonization in the outer envelope proceeds
faster than that in the inner core.
 After stellar collapse the electron lepton number is
trapped inside the matter.
\newline
The collapse of an iron core of a massive star leads to the formation
of a core residue, namely a proto-neutron star which can be
considered as an intermediate stage before the formation of a
cold  compact neutron star. After the core  collapse the electron
lepton number is trapped inside the matter. This leads to large
electron neutrino $\nu_e$ chemical potential.  Evolution of a
proto-neutron star which proceeds owing to the neutrino and photon emission
causes  the star change itself from a hot, bloated object into
a cold  compact neutron star and  can be described in terms of
separate phases starting from the moment when the star becomes
gravitationally decoupled from the expanding ejection. Through  the
events which lead up to the formation of a neutron star, two
limiting cases can be distinguished: the very beginning stage
characterized by the low entropy  density $s=1$ (in units of the
Boltzmann's constant), unshocked core with trapped neutrinos
$Y_L=0.4$ and  low density, high entropy   outer layer. The final
stage  is identified with a cold deleptonized object ($Y_{\nu}=0,
s=0$). These two distinct stages are separated by the period of
deleptonization when the neutrino fraction decreases from the
nonzero initial value ($Y_{\nu}\neq 0$) which is established by
the requirement of the fixed  total electron lepton number at
$Y_L=0.4$, to the final one characterized by $Y_{\nu}=0$.
\section{Proto-neutron star model}
The presented model of a nascent neutron star is constructed under
the assumption that the star can be divided into two main parts:
the dense core and the outer layer. As it was stated above, at the
very beginning of a proto-neutron star evolution neutrinos are
trapped on the dynamical time scale within the matter both in the
core and in the outer layer. The estimated electron lepton
fraction ($Y_L=Y_e+Y_{\mu}+Y_{\nu_e}$) at trapping,  with the value of
$\approx 0.4$ and the assumption of constant entropy, allows one to
specify the star characteristics at the conditions prevailing in
the star interior.  Assuming the evolutionary scenarios presented
in the paper by Prakash et al.\cite{Prakash} the following stages
in the life of a proto-neutron star have been distinguished:
\begin{itemize}
\item stage 1: the post bounce phase - the bounced inner core
settles into a hydrostatic configuration. Under the assumption
of  very low kinetic energy of the matter behind the shock also
the structure of the outer envelope can be approximated by
hydrostatic equilibrium. Thus a proto-neutron star model
immediately after the collapse can be constructed basing on the
following physical conditions: the low entropy core $s=1$ with
trapped neutrinos $Y_{L}=0.4$  is surrounded by high entropy
$s=2-5$ outer layer also with trapped neutrinos.
  The energy is carried out from a proto-neutron star by neutrinos and
antineutrinos of all flavors whereas lepton number is lost by the
emission of  electron neutrinos which are produced in the $\beta$
process $e^-+p\rightarrow \nu_e+n$. Deleptonization in the outer
envelope proceeds faster than that in the inner core.
\item stage 2: deleptonization of the outer layer is completed. The  physical
conditions that characterized the whole system are as follows: the
core with the entropy $s=2$ and $Y_{L}=0.4$ and the deleptonized
outer envelop with $s=2$ and $Y_{\nu}=0$.
\item stage 3: during this stage the deleptonization of the core takes place, after which  the core has neutrino-free, high entropy $s=2$ matter. Thermally produced neutrino pairs of all flavors are
abundant, and they are emitted with very similar luminosities from
the thermal bath of the core.
\item stage 4: cold, catalyzed object.
\end{itemize}
\section{The equation of state and the equilibrium conditions}
The solution of the hydrostatic equilibrium equation of Tolman,
Oppenheimer and Volkov which allows one to construct a theoretical
model of a proto-neutron star and to specify and analyze its
properties demands the specification of the equation of state. The
mean field approach to the relativistic theory of hadrons has been
used  extensively in order to describe properties of nuclear
matter and finite nuclei. The degrees of freedom relevant to this
theory are nucleons interacting through the exchange of
isoscalar-scalar $\sigma$, isoscalar-vector $\omega$,
isovector-vector $\rho$ and the pseudo-scalar $\pi$ mesons. In the
relativistic mean field approach nucleons are considered as Dirac
quasiparticles moving in classical meson fields. The contribution
of the $\pi$ meson vanishes at the mean field level. The chiral
effective Lagrangian proposed by Furnstahl, Serot and Tang
constructed on the basis of effective field theory and density
functional theory for hadrons gave in the result the extension of
the standard relativistic mean field theory and introduced
additional non-linear scalar-vector and vector-vector
self-interactions. This Lagrangian in general includes all
non-renormalizable couplings consistent with the underlying
symmetries of QCD and can be considered as the one of the
effective field theory of low energy QCD. Applying the dimensional
analysis of Georgi and Manohar \cite{georgi1}, \cite{georgi2} and
the concept of naturalness one can expand the nonlinear Lagrangian
and organize it in increasing powers of the fields and their
derivatives and truncated at given level of accuracy \cite{serot}.
 In the high density regime in neutron star interiors when the Fermi
energy of nucleons exceeds the hyperon masses additional hadronic
states are produced \cite{glen}, \cite{glen1}, \cite{bema},
\cite{weber}, \cite{gal}.
Thus the considered model involves the full octet of baryons interacting through the exchange of
$\sigma$ mesons which produce the medium range attraction and the exchange of $\omega$
mesons responsible for the short range repulsion. The model also
includes the isovector mesons $\rho $. In order to reproduce
attractive hyperon-hyperon interaction two additional
hidden-strangeness mesons, which do not couple  to nucleons, have
been introduced, namely the scalar meson $f_0(975)$
(denoted as $\sigma^{\ast}$) and the vector meson $\phi(1020)$ \cite{Schaffner}.
If the truncated Lagrangian includes terms up to the forth order
it can be written in the following form:
\begin{widetext}
\begin{eqnarray}
\mathcal{L} & = & \sum_{B}
\bar{\psi}_B(i\gamma^{\mu}D_{\mu}-m_B+g_{\sigma
B}\sigma+g_{\sigma^{\ast}B}\sigma^{\ast})\psi_B \\ \nonumber &+&
\frac{1}{2}\partial _{\mu }\sigma \partial ^{\mu }\sigma -m_{\sigma}^2\sigma^2(\frac{1}{2}+\frac{\kappa_3}{3!}\frac{g_{\sigma B}\sigma}{M}+\frac{\kappa_4}{4!}\frac{g_{\sigma B}^2\sigma^2}{M^2})+ \frac{1}{2}\partial _{\mu }\sigma^* \partial ^{\mu }\sigma^*- \frac{1}{2}m^{2}_{\sigma^*}\sigma^{*2}+ \nonumber \\
 & + &\frac{1}{2}m^{2}_{\phi}\phi_{\mu}\phi ^{\mu} -\frac{1}{4}\phi_{\mu \nu}\phi^{\mu \nu}
 -\frac{1}{4}\Omega _{\mu \nu }\Omega ^{\mu \nu }+\frac{1}{2}(1+\eta_{1}\frac{g_{\sigma B}\sigma}{M}+\frac{\eta_2}{2}\frac{g_{\sigma B}^2\sigma^2}{M^2})m_{\omega }^2\omega _{\mu }\omega ^{\mu }+ \nonumber \\
 & - & \frac{1}{4}R_{\mu \nu }^{a}R^{a\mu \nu }+(1+\eta_{\rho}\frac{g_{\sigma B}\sigma}{M})\frac{1}{2}m_{\rho} ^2\rho^{a}_{\mu }\rho^{a\mu }+\frac{1}{24}\zeta_0 g_{\omega B}^2(\omega _{\mu }\omega ^{\mu })^{2}. \label{lag1}
\end{eqnarray}
where $\Psi_B^T
=(\psi_N,\psi_{\Lambda},\psi_{\Sigma},\psi_{\Xi})$.  The covariant
derivative $D_{\mu}$ is defined as
\begin{equation}
D_{\mu}=\partial_{\mu}+ig_{\omega B}\omega_{\mu}+ig_{\phi
B}\phi_{\mu}+ig_{\rho B}I_{3B}\tau^a\rho^a_{\mu}
\end{equation}
whereas  \( R_{\mu \nu }^{a} \), \( \Omega _{\mu \nu } \) and \(
\phi_{\mu \nu } \) are the field tensors
\begin{equation}
R_{\mu \nu }^{a}=\partial _{\mu }\rho^{a}_{\nu }-\partial _{\nu
}\rho^{a}_{\mu }+g_{\rho }\varepsilon _{abc}\rho_{\mu
}^{b}\rho_{\nu }^{c},
\end{equation}
\begin{equation}
\Omega _{\mu \nu }=\partial _{\mu }\omega _{\nu }-\partial _{\nu
}\omega _{\mu },\hspace{2cm}\phi_{\mu \nu }=\partial _{\mu }\phi
_{\nu }-\partial _{\nu }\phi _{\mu },
\end{equation}
$m_B$ denotes baryon mass whereas \( m_{i} \) $(i= \sigma ,\omega
,\rho ,\sigma^*,\phi )$ are masses assigned to the meson fields,
$M$ is the nucleon mass. There are two parameter sets presented in
the original paper by Furnstahl et al \cite{FST1}, \cite{FST2} G1
and G2 which have been determined by calculating nuclear
properties such as binding energies, charge distribution and
spin-orbit splitting for a selected set of nuclei \cite{sil}.
 For the purposes of this paper the G2 parameter
set  has been chosen.  Calculations performed with the use of this
parameter set  properly reproduce properties of finite nuclei and
make it possible to compare the obtained results with the
Dirac-Brueckner-Hartree-Fock (DBHF) calculations for nuclear and
neutron matter above the saturation density. The DBHF method
results in a rather soft equation of state in the vicinity of the
saturation point and  for higher densities. Calculations performed
on the basis of the effective FST Lagrangian with the use of the
G2 parameter sets
predict similar, soft equation of state.
 Due to the fact that the
expectation value of the $\rho$ meson field is an order of
magnitude smaller than that of $\omega$ meson field,  the
Lagrangian function (\ref{lag1}) does not include the quartic
$\rho$ meson term. In addition, as this paper deals with the
problem of infinite nuclear matter the terms in the Lagrangian
function in the paper \cite{FST2} involving tensor couplings and
meson field gradients have been excluded.
%\newline
The derived equations of motion constitute a set of coupled
equations which have been solved in the mean field approximation.
In this approximation meson fields are separated into classical
mean field values and quantum fluctuations which are not included
in the ground state
\begin{center}
\begin{tabular}{lll}
$\sigma  = \overline{\sigma}$ +$ s_0$ &$ \sigma^*  =
\overline{\sigma}^*$
+ $s^*_0$ &   \\
$\phi_{\mu}  = \overline{\phi}_{\mu} + f_0\delta_{\mu 0}$ &
$\omega_{\mu} = \overline{\omega}_{\mu}+ w_{0}\delta_{\mu 0}$ &
$\rho_{\mu}^a  =  \overline{\rho}^a_{\mu}+r_{0}\delta_{\mu
0}\delta^{3a}$.
\end{tabular}
\end{center}
\vspace{0.5cm}
 As it was stated above the
derivative terms are neglected and only time-like components of
the vector mesons will survive if one assumes homogenous and
isotropic infinite matter. The field equations derived from the
Lagrange function at the mean field level are
\begin{equation}
m_{\sigma}^2(s_0+\frac{g_{\sigma
B}\kappa_3}{2M}s_0^2+\frac{g_{\sigma B}^2\kappa_4}{6
M^2}s_0^3)-\frac{1}{2}m_{\omega}^2(\eta_1\frac{g_{\sigma
B}}{M}+\eta_2\frac{g_{\sigma B}^2}{M^2}s_0
)w_0^2-\frac{1}{2}m_{\rho}^2\eta_{\rho}\frac{g_{\sigma}}{M}r_0^2=\sum_Bg_{\sigma
B}m^2_{eff,B}S(m_{eff,B})
\end{equation}
\begin{equation}
m_{\omega}^2(1+\frac{\eta_1g_{\sigma}}{M}s_0+\frac{\eta_2g_{\sigma}^2}{2M^2}s_0^2)w_{0}+
\frac{1}{6}\zeta_0g_{\omega B}^2w_{0}^3=\sum_Bg_{\omega B}n_B
\end{equation}
\begin{equation}
m_{\rho}^2(1+\frac{g_{\sigma
B}\eta_{\rho}}{M}s_0)r_0=\sum_Bg_{\rho B}I_{3B}n_B
\end{equation}
\begin{equation}
m_{\sigma*}^2s_0^*=\sum_Bg_{\sigma^*B}m^2_{eff,B}S(m_{eff,B})
\end{equation}
\begin{equation}
m_{\phi}^2f_0=\sum_Bg_{\phi B}n_B.
\end{equation}
The function $S(m_{eff,B})$ is expressed with the use of the
integral
\begin{equation}
S(m_{eff,B})=\frac{2J_B+1}{2\pi^2}\int_0^\infty
\frac{k^2dk}{E(k,m_{eff,B})}(f_B-f_{\overline{B}})
\end{equation}
where $J_B$ and $I_{3B}$ are the spin and isospin projection of
baryon $B$,  $n_B$ is given by
\begin{equation}
n_B=\frac{2J_B+1}{2\pi^2}\int_0^\infty k^2dk(f_B-f_{\bar{B}})
\end{equation}
 The functions $ f_B$ and $f_{\bar{B}} $ are the
Fermi-Dirac distribution for particles and anti-particles,
respectively
\begin{equation}
f_{B,\bar{B}}=\frac{1}{1+e^{E[(k,m_{eff,B})\mp \mu_B]/k_BT}}.
\end{equation}
where $\mu_B$ denotes the baryon chemical potential defined as
\begin{equation}
\mu_B=\nu_B+g_{\omega B}w_0+g_{\rho B}I_{3B}r_0+g_{\phi
B}f_0=\sqrt{k^2+m_{eff,B}^2}+g_{\omega B}w_0+g_{\rho
B}I_{3B}r_0+g_{\phi B}f_0 \label{potchem}
\end{equation}
 The obtained Dirac equation for baryons has the following form
\begin{equation}
(i\gamma ^{\mu }\partial_{\mu }-m_{eff,B}-g_{\omega
B}\gamma^{0}\omega_{0}-g_{\phi B}\gamma^{0}f_{0})\psi_B =0
\end{equation}
with $m_{eff,B}$ being the effective baryon mass generated by the
baryon and scalar fields interaction and defined as
\begin{equation}
m_{eff,B}=m_B-(g_{\sigma B}s_0+g_{\sigma^*B}s_0^*).
\label{masseff}
\end{equation}
In the case of a proto-neutron star matter which includes hyperon
degrees of freedom the considered parameterization  has to be
supplemented by the parameter set related to the strength of the
hyperon-nucleon and hyperon-hyperon interactions. The scalar meson
coupling to hyperons can be calculated from the potential depth of
a hyperon in the saturated nuclear matter \cite{Muller}
\begin{equation}
U^{N}_{Y}=-g_{\sigma Y}s_0 +g_{\omega Y}\omega_{0}
\end{equation}
 The considered model does not include
$\Sigma$ hyperons due to the remaining uncertainty of the form of
their potential in nuclear matter at saturation density
\cite{Mares}, \cite{lamsig}, \cite{ksi}, \cite{ksi2}. In the
scalar sector the scalar coupling of the $\Lambda$ and $\Xi$
hyperons requires constraining in order to reproduce the estimated
values of the potential felt by a single $\Lambda$ and a single
$\Xi$ in saturated nuclear matter ($\rho_0 \sim 2.5 \ 10^{14} \  g/cm^3 $)
\begin{eqnarray}
U^{(N)}_{\Lambda}(\rho_0)&=&g_{\sigma
\Lambda}s_0(\rho_0)-g_{\omega \Lambda}w_0(\rho_0)\simeq 27-28 MeV\\
\nonumber
 U^{(N)}_{\Xi}(\rho_0)&=&g_{\sigma
\Xi}s_0(\rho_0)-g_{\omega \Xi}w_0(\rho_0)\simeq 18-20 MeV.
\end{eqnarray}
Assuming the SU(6) symmetry for the vector coupling constants and
determining the scalar coupling constants from the
potential depths, the hyperon-meson couplings can be fixed.\\
 The strength of hyperon coupling
to strange meson $\sigma^{\ast}$ is restricted through the
following relation \cite{gal}
\begin{equation}
U^{(\Xi)}_{\Xi}\approx U^{(\Xi)}_{\Lambda}\approx
2U^{(\lambda)}_{\Xi}\approx 2U^{(\lambda)}_{\Lambda}.
\end{equation}
which together with the estimated value of hyperon potential
depths in hyperon matter provides effective constraints  on scalar
coupling constants to the $\sigma^{\ast}$ meson. The currently
obtained value of the $U^{(\Lambda)}_{\Lambda}$ potential at the
level of 5 MeV \cite{Takahashi} permits the existence of
additional parameter set \cite{bema2} which reproduces this weaker
$\Lambda\Lambda$ interaction. In the text this parameter set is
marked as weak, whereas strong denotes the stronger
$\Lambda\Lambda$ interaction for $U^{(\Lambda)}_{\Lambda}\simeq
20$ MeV \cite{Schaffner}.
 The vector coupling constants for hyperons
which are determined from $SU(6)$ symmetry constraints \cite{shen}
remain unchanged. They are of primary importance in determining
the high density range of repulsive baryon potentials \cite{shen}
as
\begin{eqnarray}
\frac{1}{2}g_{\omega \Lambda}=\frac{1}{2}g_{\omega
\Sigma}=g_{\omega \Xi}=\frac{1}{3}g_{\omega N}  \\ \nonumber
\frac{1}{2}g_{\rho \Sigma}=g_{\rho \Xi}=g_{\rho N}; g_{\rho
\Lambda}=0  \\ \nonumber 2g_{\phi \Lambda}=2g_{\phi
\Sigma}=g_{\phi \Xi}=\frac{2\sqrt{2}}{3}g_{\omega N}.
\end{eqnarray}
\begin{table}
\caption{Chosen parameter sets.}\label{tab:TM1}
\large{
\begin{center}
\begin{tabular}{lllllllll}
  \hline
&$m_{\sigma}$[MeV]&$g_{\sigma}$&$g_{\omega}$&$g_{\rho}$&$\eta_1$&$\eta_2$&$\eta_{\rho}$&$\zeta_0$\\
\hline
G2&520.254&10.4957&12.7624&9.48346&0.64992&0.10975&0.3901&2.6416\\\hline
\end{tabular}
\end{center}
}
\end{table}
\begin{table}
\caption{Strange scalar sector parameters.}\label{tab:sscalar}
\large{
\begin{center}
\begin{tabular}{ll|l|l|l|l}
  \hline
&&$g_{\sigma\Lambda}$&$g_{\sigma\Xi}$&$g_{\sigma^{\ast}\Lambda}$&$g_{\sigma^{\ast}\Xi}$\\
\hline G2&weak&6.41007&3.33730&3.89005&11.64347\\\cline{2-6} &strong&6.41007&3.33730&7.93120&12.45836\\
\hline
\end{tabular}
\end{center}
}
\end{table}
Since the dynamical time scale of  neutron star evolution is much
longer than the weak decay time scale of baryons (including
strangeness violating processes) neutron star matter is considered
as a system with conserved baryon number $n_b=\sum_B n_B$
($B=n,p,\Lambda,\Xi^-, \Xi^0$) and electric charge being in
chemical equilibrium with respect to weak decays. The chemical
equilibrium is characterized by the correlations between chemical
potentials which in the case of the process $p+e^-\leftrightarrow
n+\nu_e$ leads to the following relation
\begin{equation}
\mu_{asym} \equiv \mu_n-\mu_p =\mu_e-\mu_{\nu_{e}}.
\end{equation}
The similar weak process  $\mu+\nu_{e}\leftrightarrow
e+\nu_{\mu}$ leads to
\begin{equation}
\mu_{\mu} = \mu_e-\mu_{\nu_{e}}+\mu_{\nu_{\mu}} =\mu_{asym}+\mu_{\nu_{\mu}}.
\end{equation}
Assuming that the muon neutrinos are not trapped  inside the protoneutron star ($\mu_{\nu_{\mu}}=0$) these relations involve three independent chemical potentials
$\mu_n$, $\mu_e$ and $\mu_{\nu_{e}}$ corresponding to baryon
number, electric charge and lepton number conservation. In general,
weak processes for baryons can be written in the following form
\begin{equation}
B_1+l\leftrightarrow B_2+\nu_{l}
\end{equation}
where $B_1$ and $B_2$ are baryons, $l$ and $\nu_l$  denote lepton
and neutrino of the corresponding flavor, respectively. Provided
that the  weak processes stated above  take place in thermodynamic
equilibrium the following relation between chemical potentials
can be established
\begin{equation}
\mu_B=q_B\mu_n-q_{e_{B}}(\mu_e-\mu_{\nu_e}). \label{potchem1}
\end{equation}
In this equation $\mu_B$ denotes chemical potential of baryon $B$
with the baryon number $q_B$ and the electric charge $q_{e_{B}}$.
Using the relation (\ref{potchem1}) for $\Lambda$, $\Sigma$ and
$\Xi$ hyperons the following results can be obtained
\begin{equation}
\mu_{\Lambda}=\mu_{\Sigma^{0}}=\mu{\Xi^{0}}=\mu_n,
\hspace{5mm}\mu_{\Sigma^{-}}=\mu_{\Xi_{-}}=\mu_n+(\mu_e-\mu_{\nu_{e}}),
\hspace{5mm}\mu_p=\mu_{\Sigma^{+}}=\mu_{n}-(\mu_{e}-\mu_{\nu_{e}}).
\end{equation}
 The presented constrains of charge neutrality and
 $\beta$-equilibrium
imply the presence of leptons which are introduced as free
particles. Thus the Lagrangian of free leptons has to be added to
the Lagrangian function (\ref{lag1})
\begin{equation}
\mathcal{L}_{L}=\sum_{i=e,\mu}\overline{\psi}_{i}(i\gamma ^{\mu
}\partial_{\mu }-m_{i})\psi_{L}+\sum_{i=e,\mu}\overline{\psi}_{\nu_{i}}(i\gamma ^{\mu
}\partial_{\mu })\psi_{\nu_{i}}.
\end{equation}
\end{widetext}
Muons start to appear in neutron star matter in the process
$e^-\leftrightarrow \mu^-+\bar{\nu}_{\mu}+\nu_{e}$ after $\mu_{\mu}$ has
reached the value equal to the muon mass. The appearance of muons
not only reduces the number of  electrons but also affects the
proton fraction. Numerical models of the collapsing inner core
have shown that there are no muon neutrinos ($\mu_{\nu_{\mu}}=0 $) and
only electron neutrinos are trapped in the matter  \cite{Prakash}.
 After the deleptonization of a proto-neutron star matter  there are no
trapped neutrinos either and $\mu_{\nu_{e}}= 0$. In this case the number
of independent chemical potentials reduces to two ($\mu_n$ and
$\mu_e$).
%\end{widetext}
\section{Results}
More realistic description of a neutron star requires taking into consideration
not only the interior region of a neutron star but also the remaining
layers, namely the inner and outer crust and surface layers. The
composite equation of state has been constructed by adding Bonn
and Negele-Vautherin (NV) equations of state (describing the inner
crust) to the  one determined with the use of the G2
parameterization of the FST model (\cite{FST1},\cite{SerotI})
allows to calculate the neutron star structure
for the entire neutron star density span. The proto-neutron star
models have been constructed on the assumption that they include a
core and an outer layer. The outer region of a proto-neutron star
is characterized by the following conditions: it is less dense and
less massive than the core, the assumptions of nearly complete
disintegration of nuclei into free nucleons and $\beta$
equilibrium have been made. Chemical potentials of the
constituents of the matter are evaluated by the requirement of
charge neutrality and $\beta$ equilibrium.   At sufficiently low
density, in the absence of interactions which alter the nucleon
effective mass nucleons are considered as non-relativistic. The
pressure in the outermost layer of a proto-neutron star is
determined by nucleons, electrons and neutrinos. The theoretical
model of the dense inner core can be constructed  under the
following assumptions:  the matter includes the full octet of
baryons interacting through the exchange of meson fields, the
composition is determined by the requirements of charge neutrality
and generalized $\beta$ equilibrium.
 The loss  of lepton number from the collapsed star
proceeds in separated stages. First, very fast, carried on in a
very short time in comparison with the Kelvin-Helmholtz neutrino
cooling time, the deleptonization of the surface layer takes place
\cite{Prakash}. After that the deleptonization of the core
proceeds by the emission of electron neutrinos which are produced
efficiently via the $\beta$-process
\begin{equation}
e^{-}+p \rightarrow n+\nu_{e}.
\end{equation}
\begin{widetext}
The events between the two distinguished  moments depend on
details of the considered models, especially on the equation of
state.  The initial moment $t=0$ characterizes the stage with
neutrinos trapped  $(Y_{L}=0.4)$ both in the core and in the outer
envelope. Introducing a parameter $\alpha$, which determines the
degree of deleptonization of the proto-neutron star matter, the
moment $t=0$ can be  identified with $\alpha=1$ whereas the final
completely  deleptonized stage $(Y_{\nu}=0)$ is described by
$\alpha=0$. The density at the core-outer layer interface is time
dependent. This dependance can be presented by comparison of the
interface between the inner core equation of state and the
equations of state which have been obtained for different values
of the parameter $\alpha$ in the outer layer. Results are shown in
Fig. \ref{coreenvelop}. The deleptonization leads to the
substantial reduction in the extension of the proto-neutron star
envelope.
\begin{figure}
\subfigure {\includegraphics[  width=8.15cm]{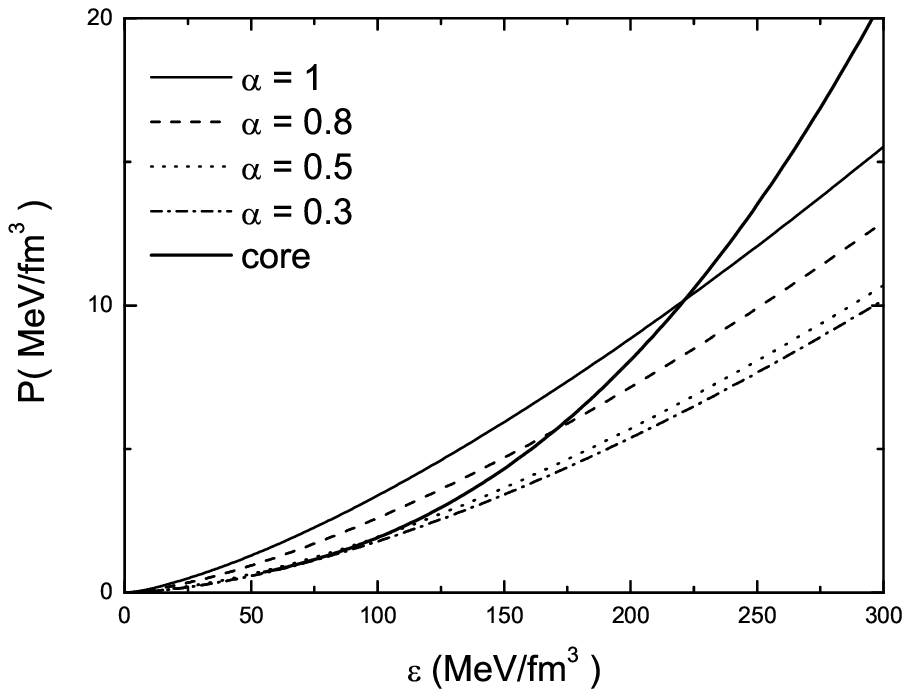}} \subfigure
{\includegraphics[  width=8.15cm]{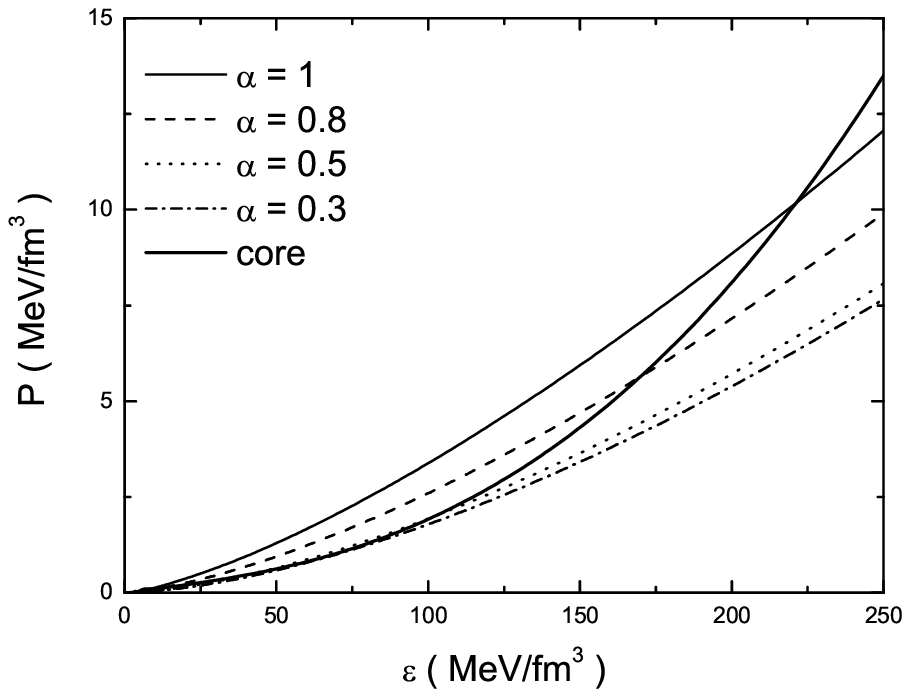}} \caption{The density
dependance of the interface  between the inner core equation of
state and the equations of state which have been obtained for
different values of the parameter $\alpha$ in the outer layer for
the  the strong (left panel) and weak (right panel) $Y-Y$
interaction models.} \label{coreenvelop}
\end{figure}
 The obtained form of the
equation of state for the strong and weak $Y-Y$ interactions are
presented in Fig. \ref{fig:eos}.
\begin{figure}
\subfigure {\includegraphics[width=8.15cm]{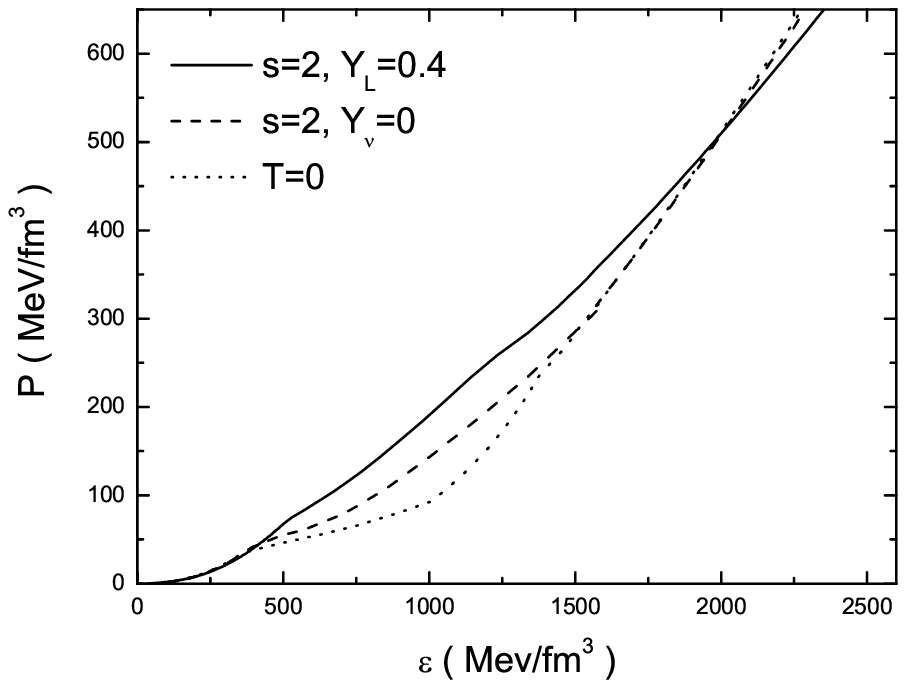}}
\subfigure {\includegraphics[width=8.15cm]{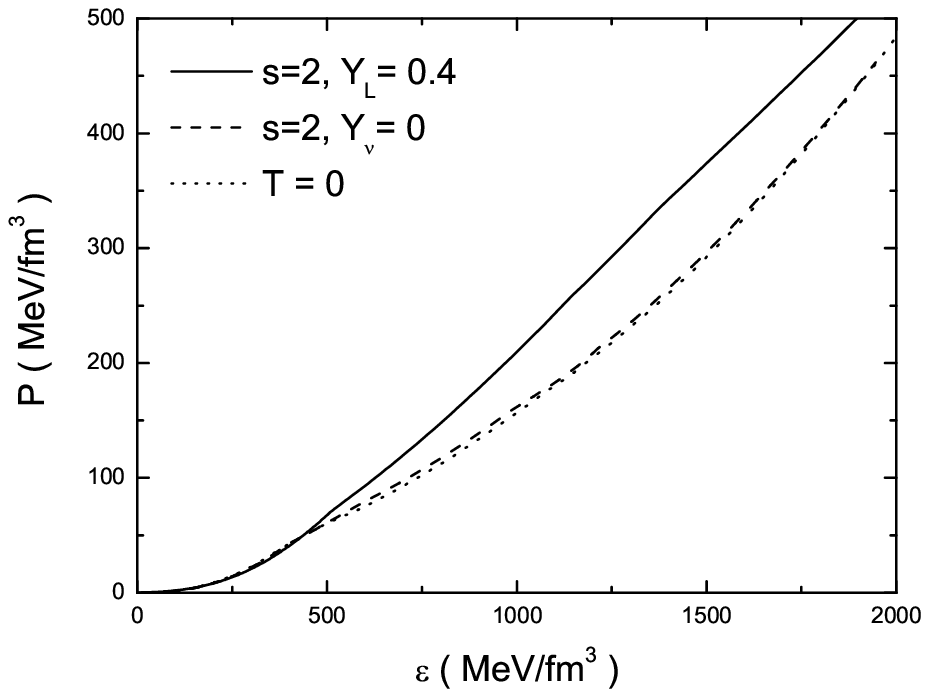}}
\caption{The equation of state of the hyperon star
matter for the G2 parameter set. In the left and right panels the
results for the strong and weak $Y-Y$ interactions  are
presented.} \label{fig:eos}
\end{figure}
Both models have yielded identical results up to a certain value
of density: the hot, neutrino-trapped matter is described by the
stiffest equation of state whereas the hot, deleptonized and cold
($T=0$)  proto-neutron star models result in softer equations of
state. However, the strong $Y-Y$ interaction model gives for the
hot, deleptonized and for the cold $T=0$ proto-neutron star matter
equations of state stiffer than that for the weak model.
\newline
The composition of hyperon star matter as well as the the
threshold density for hyperons  are altered when the strength of
the hyperon-hyperon interaction is changed. Fig. \ref{YSnB}
presents fractions of particular strange baryon species $Y_B$ as a
function of baryon number density $n_b$ for the weak and strong
model of the G2 parameterization, for the three selected
evolutionary phases. All calculations have been done on the
assumption that the repulsive $\Sigma$ interaction shifts the
onset points of $\Sigma$ hyperons to very high densities and they
do not appear in the considered proto-neutron and neutron star
models.
\begin{figure}
\subfigure {\includegraphics[width=8.15cm]{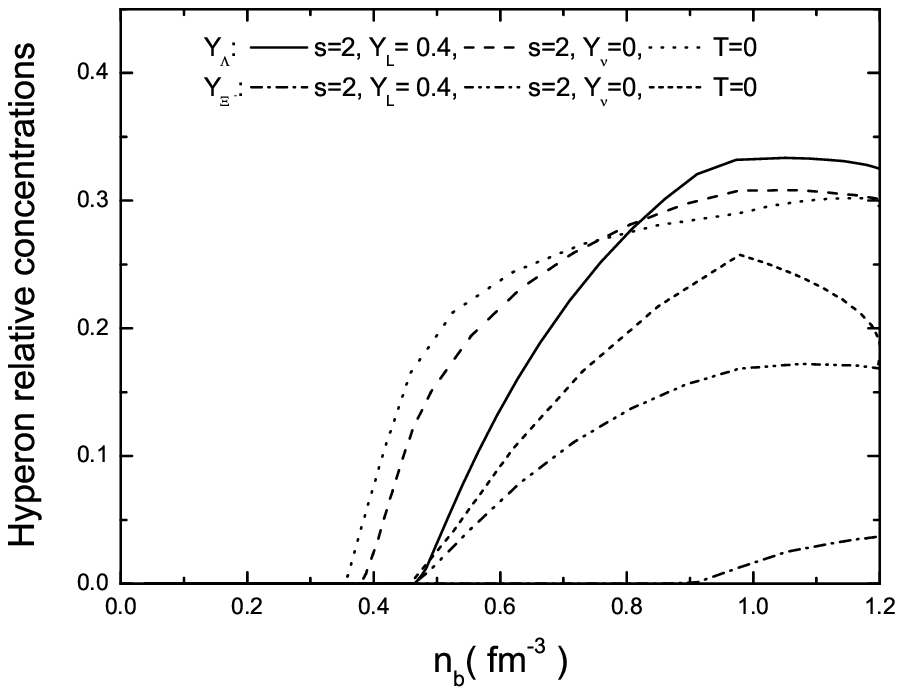}}
\subfigure {\includegraphics[width=8.15cm]{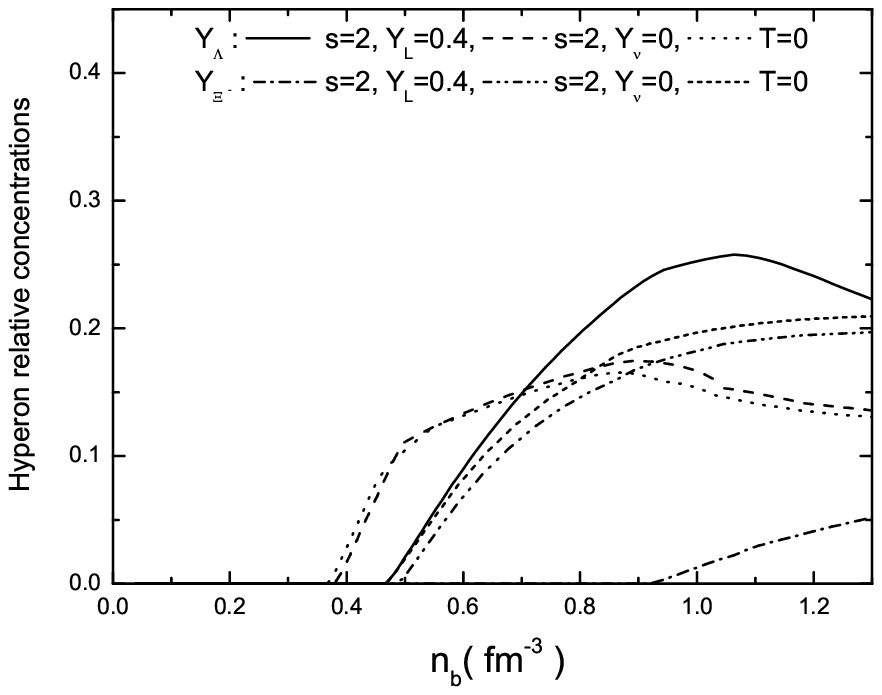}}
\caption{Relative concentrations of $\Lambda$ and $\Xi$ hyperons
in  hyperon star matter for the G2 parameter set.  In the left and
right panels the results for the strong and weak Y-Y interaction
are presented.} \label{YSnB}
\end{figure}
  The onset of
$\Xi^0$ hyperons takes place at high densities and they are not
presented in  Fig. \ref{YSnB}. In the case of strong and weak
$Y-Y$ interactions the onset of $\Lambda$ hyperons is followed by
the onset of $\Xi^-$ hyperons. In general, when $\Lambda$ hyperons
appear they replace protons and, in the consequence, lower the
energy of the system. This fact, through the requirement of charge
neutrality of neutron star matter, results in diminished electron
chemical potential. The populations of $\Lambda$ hyperons obtained
in the weak models in the considered evolutionary stages are
reduced in comparison with those calculated for the strong $Y-Y$
interaction models.  Thus the weak model  with the reduced
$\Lambda$ hyperon population leads to the increased value of
electron and proton concentrations. Results are presented in Fig.
\ref{Ye}  where relative concentrations of electrons and protons
are depicted.
\begin{figure}
\subfigure {\includegraphics[width=8.15cm]{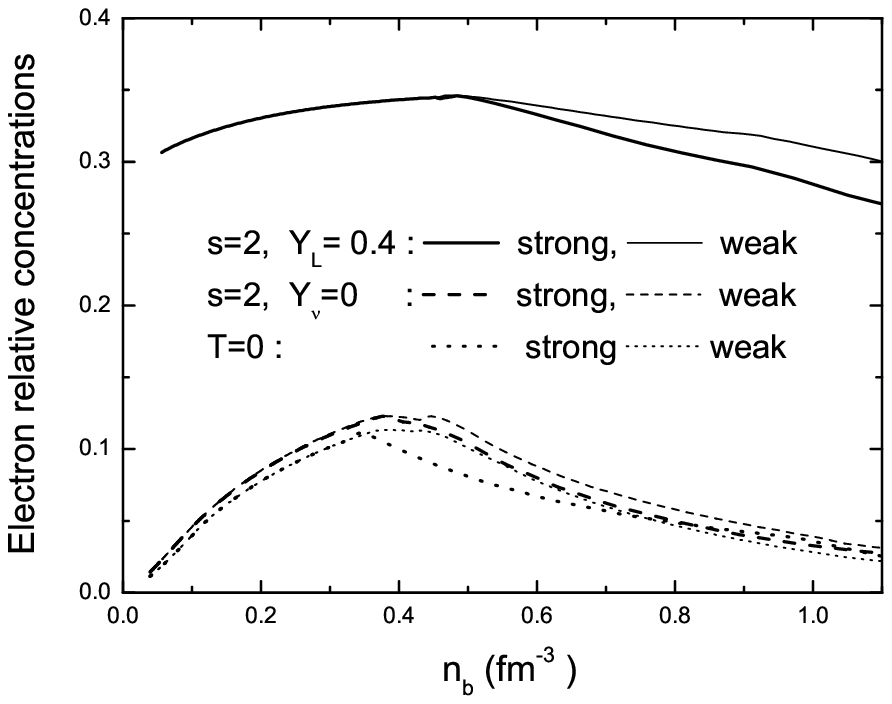}} \subfigure
{\includegraphics[width=8.15cm]{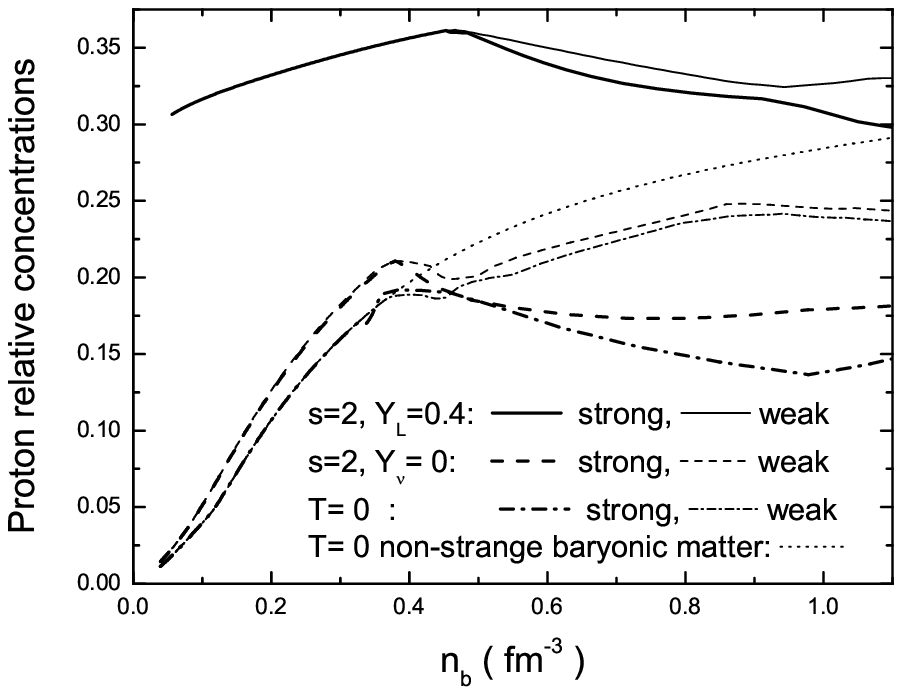}} \caption{Relative
concentrations of electrons (the left panel) and protons (the
right panel) in hyperon star matter for the G2 parameter set.}
\label{Ye}
\end{figure}
One can also compare the concentrations of neutrinos in  the hot,
neutrino trapped matter. In Fig. \ref{Fig:Ynu}  neutrino
concentrations for the two different models are shown.
\begin{figure}
\subfigure {\includegraphics[width=8.15cm]{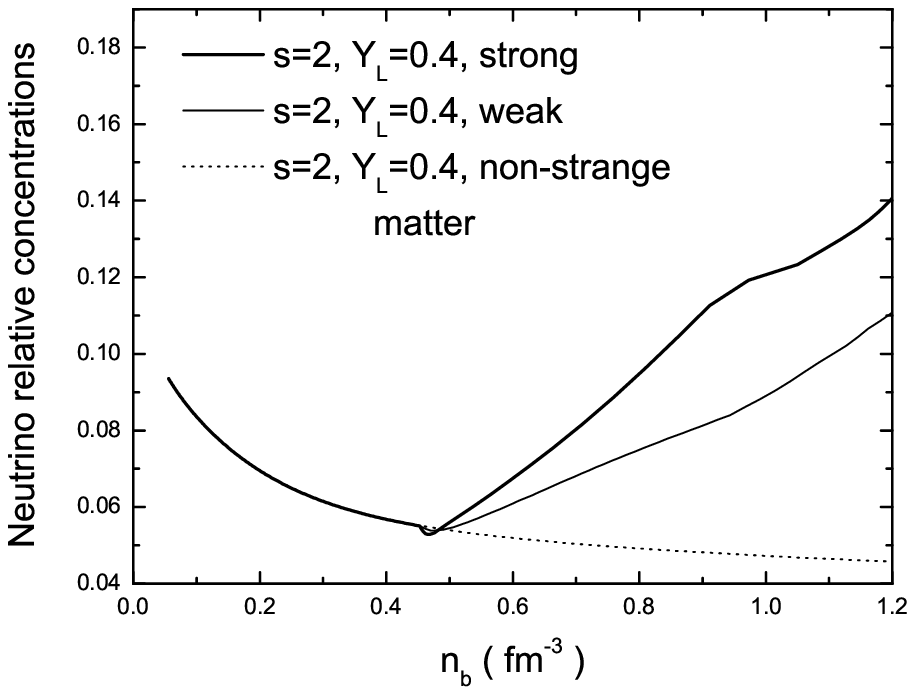}} \subfigure
{\includegraphics[width=8.15cm]{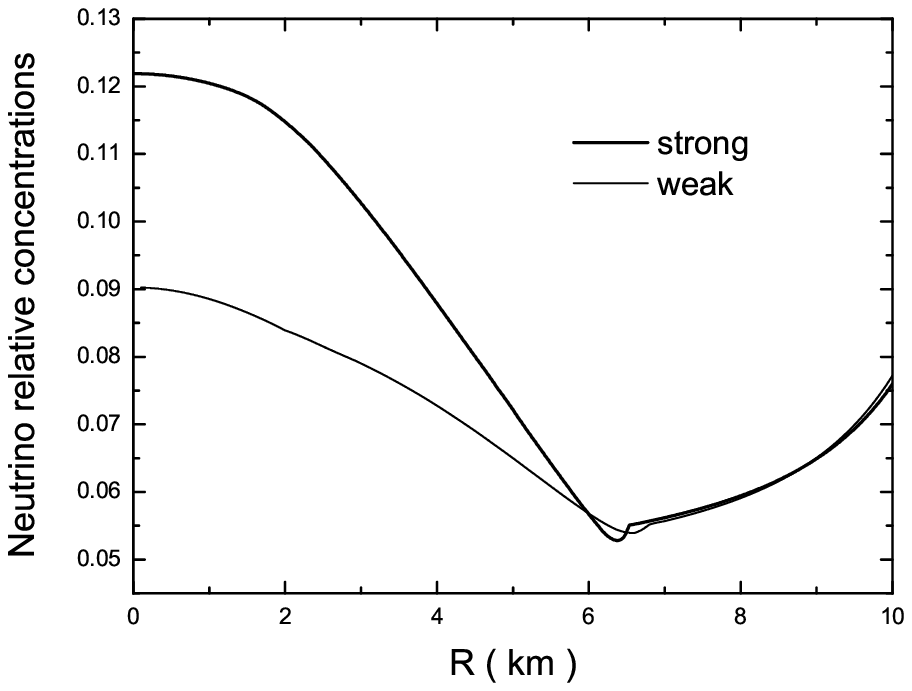}} \caption{Left panel:
Density dependance of neutrino concentrations. Right panel:
Neutrino concentrations as a function of stellar radius R.}
\label{Fig:Ynu}
\end{figure}
 The weak model leads to lower
concentrations of neutrinos.  For comparison results for
non-strange proto-neutron star matter is included.
\newline
Once the equations of state have been calculated for each
evolutionary phase and for the strong and weak models the
corresponding hydrostatic models of proto-neutron stars have been
obtained.
 Solutions of the Oppenheimer-Tolman-Volkov equation for the
considered parameter sets are presented in Fig. \ref{fig:rm}.
\begin{figure}
\subfigure {\includegraphics[width=8.15cm]{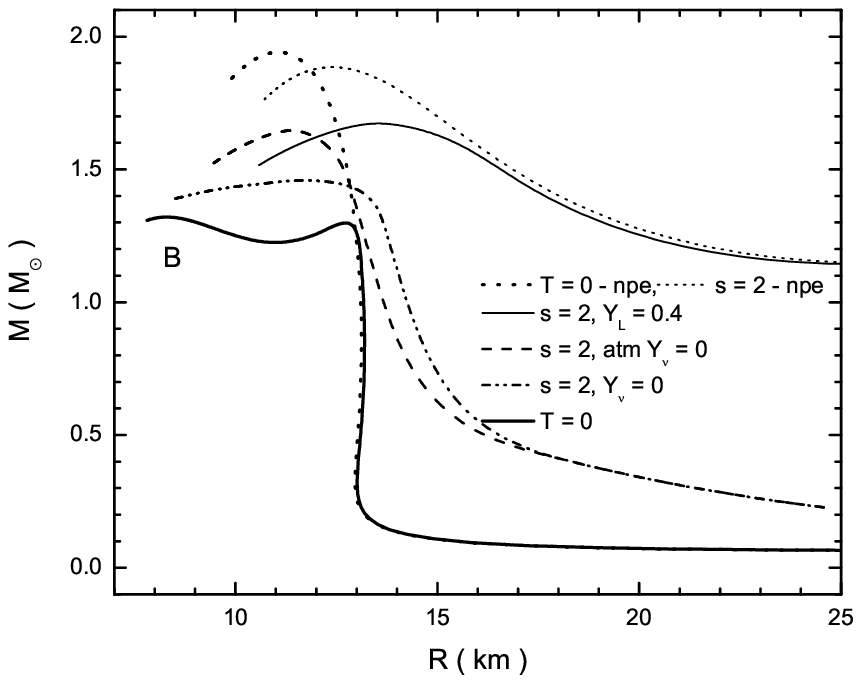}}
\subfigure{\includegraphics[width=8.15cm]{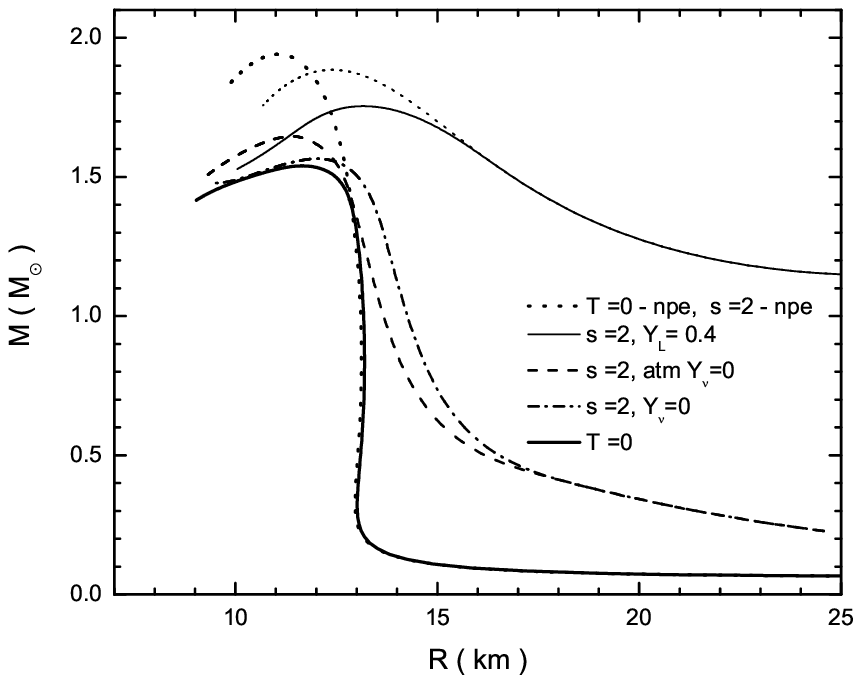}}
\caption{The mass-radius relations for subsequent stages of the proto-neutron
star evolution. The left panel presents results for G2
parameterization with the strong $Y-Y$ interaction, the right
panel for the weak interaction model. Dotted lines represent
solutions obtained for non-strange baryonic matter.}
\label{fig:rm}
\end{figure}
In agreement with the introduced evolutionary stages of a
proto-neutron star the obtained mass-radius relations include a
sequence of models calculated for a given equation of state
characteristic for a specific phase of a proto-neutron star
evolution. In the left and right panel of Fig. \ref{fig:rm}
results for six equations of state have been plotted. The four
mass-radius relations are constructed for the strangeness-rich
baryonic matter and they represent the evolution of a
proto-neutron star starting from the moment when the proto-neutron
star can be modelled as a low entropy  core surrounded by a high
entropy outer layer. Neutrinos are trapped both  in  the core and
in the outer layer. This very beginning phase of a proto-neutron
star is followed by two subsequent stages connected with the
process of deleptonization of a proto-neutron star. The first
during which the deleptonization of the outer layer takes place
whereas in the core neutrinos are still trapped and the second one
representing a hot deleptonized object with thermally produced
neutrino pairs of all flavor abundant in the core and in the outer
layer. The final case is exemplified by a solution obtained for
cold neutron star matter which includes hyperons. For comparison
the mass-radius relations for two limiting cases of  proto-neutron
star evolution for nonstrange baryonic matter have been presented.
A similar sequence of the mass-radius relations have been
constructed for the G2 parameterization supplemented by the weak
$Y-Y$ parameter set for the strange sector. There are qualitative
changes in proto-neutron star parameters which occur  for a given
evolutionary stage. The G2-strong parameterization leads to
proto-neutron neutron stars with the reduced value of the maximum
mass. For the strong hyperon-hyperon interaction strength in the
case of the cold neutron star model besides the ordinary neutron
star branch there exists additional stable branch of solutions
which are characterized by a similar value of the mass but with
a significantly reduced radius. For the purpose of this paper A
denotes the maximum mass configuration of the ordinary neutron
star branch and B the additional maximum.
 Proto-neutron and neutron star parameters are
summarized in Tables \ref{tab:RMstrong} and \ref{tab:RMweak}. The
presented results have been obtained for maximum mass
configurations.
\begin{table}
\caption{Proto-neutron and neutron star parameters obtained for
the maximum mass configurations for the G2 parameterization with
the strong $Y-Y$ interactions.}
\large{
\begin{center}
\begin{tabular}{|c|c|l|l|l|l|}
\hline
 \multicolumn{2}{|c|}{}& $\rho/\rho_{0}$\,& R (km)\,& M\,$(M_{\odot})$\,&
$M_{B}\,(M_{\odot})$\,\\ \hline    T=0&A& 5.00& 12.73& 1.298& 1.407\\
\cline{2-6}&B & 21.76& 8.22& 1.320& 1.634\\\hline
\multicolumn{2}{|c|}{s=2}& 8.76& 13.528& 1.670& 1.859\\ \hline
\end{tabular}
\end{center}
}
\label{tab:RMstrong}
\end{table}
\begin{table}
\caption{Proto-neutron and neutron star parameters obtained for
the maximum mass configurations for the G2 parameterization with
the weak $Y-Y$ interactions.}
\large{
\begin{center}
\begin{tabular}{|c|c|c|c|c|}
\hline & $\rho/\rho_{0}$\,& R (km)\,& M~($M_{\odot}$)\,&
$M_{B}\,(M_{\odot})$\,\\ \hline  T=0& 8.00& 11.63& 1.527& 1.749\\
\hline s=2& 9.12& 13.14& 1.754& 1.916\\ \hline
\end{tabular}
\end{center}
}
\label{tab:RMweak}
\end{table}
The analysis of the existence and stability of the additional
branch in the mass-radius relation will be the subject of further
investigations.
\newline
 Neutron stars are purely gravitationally bound object. The gravitational
binding energy of a relativistic star is defined as the difference
between its gravitational and baryonic masses.
\begin{equation}
E_{b,g}=(M-M_{B})c^{2}
\end{equation}
where
\begin{equation}
M=4\pi \int
_{0}^{R}drr^{2}(1-\frac{2Gm(r)}{c^{2}r})^{-\frac{1}{2}}\rho (r)
\end{equation}
and $M_B=n_bm_B$ is given by the total baryon number $n_b$.  The
numerical solution of the above equation has been found for the
selected equations of state for the strong and weak interactions.
The results are shown in Fig. \ref{fig:ebin} which also includes
for comparison solutions for non-strange proto-neutron and neutron
star models. The obtained mass-radius relations for the
non-strange baryonic matter confirm the well-known fact that in
this case the maximum mass of a proto-neutron star with trapped
neutrinos is lower than that of cold deleptonized matter. This has
a consequence for the possibility of a black hole formation during
the relatively long lasting phase of deleptonization. In the case
of non-strange matter  a proto-neutron star is not able  to
achieve the unstable configuration due to deleptonization.
\begin{figure}
\subfigure {\includegraphics[width=8.15cm]{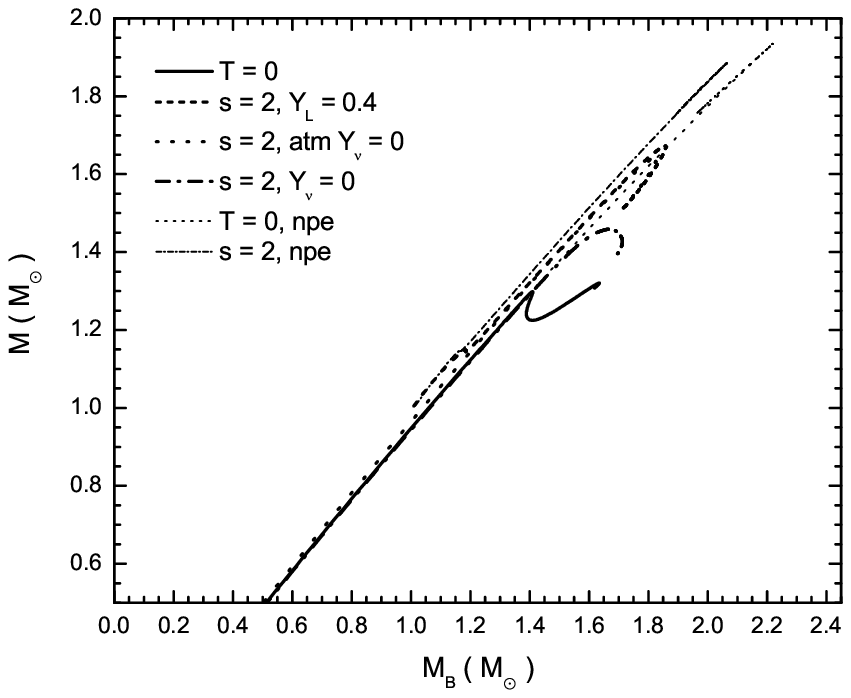}}
\subfigure {\includegraphics[width=8.15cm]{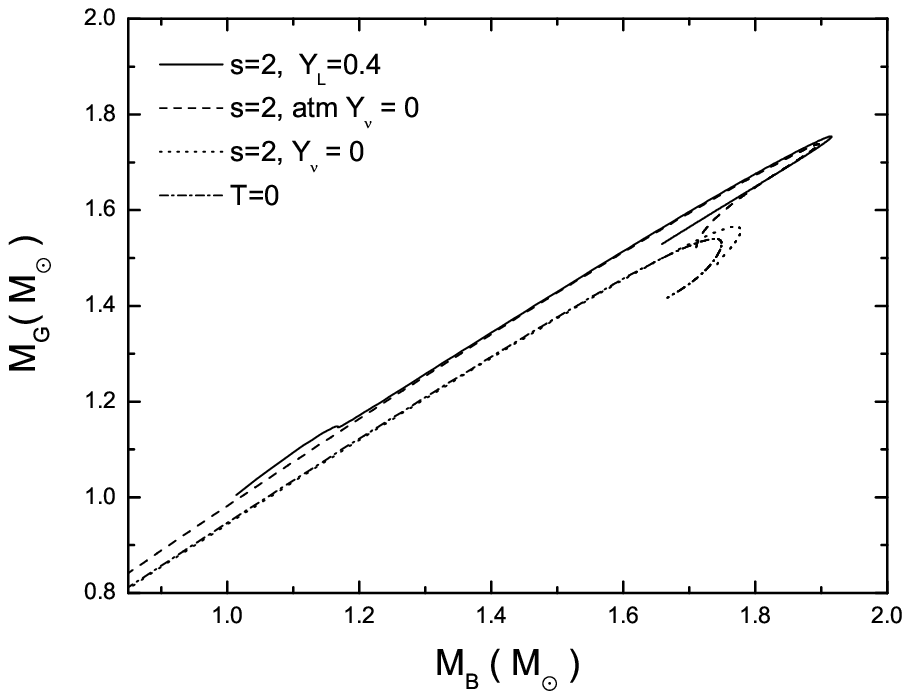}}
\caption{The neutron star mass versus the baryon mass $M_B=Mc^2 N_B$.  In the left and
right panels the results for the strong and weak Y-Y interaction,
respectively, are presented.} \label{fig:ebin}
\end{figure}
In general, neutrino trapping increases the value of the maximum
mass for the strangeness rich baryonic matter.  In all
evolutionary stages  there are configurations which due to
deleptonization go to the unstable branch of  proto-neutron and
neutron stars.
\newline
The gained solutions of the structure equations allows as to carry
out a similar analysis of the onset point,  abundance and
distributions of the individual baryon and of lepton species  but
now as functions of the star radius $R$. The analysis includes
results obtained for two extreme evolutionary phases: the very
beginning hot, neutrino-trapped matter stage and the cold,
deleptonized one. Two characteristic configurations have been
considered for the $T=0$ solution. Namely, the one connected with
the maximum mass configurations  marked as A and the other for the
B maximum mass configuration of the G2 strong parameterization.
The compact hyperon core which emerges in the interior of the
maximum mass configuration consist of $\Lambda$, $\Xi^-$ and
$\Xi^0$ hyperons. The hyperon population is reduced to $\Lambda$
and $\Xi^-$ in the case of hot neutrino trapped matter for the
weak model. The process of deleptonization and cooling diminishes
the concentration of $\Lambda$ hyperons and increases
concentrations of $\Xi^-$, $\Xi^0$ hyperons. The reduction of
$\Lambda$ hyperon population in the weak model is larger then that
obtained for the strong $Y-Y$ interaction model. The innermost
hyperon core has larger radius and contains much more negatively
charged $\Xi$ hyperons in the weak model. The relative fractions
of hyperons in the core of the maximum mass configurations are
presented in Fig. \ref{fig:xaYl} and Fig. \ref{fig:YSdweak}.
\begin{figure}
\begin{center}
{\includegraphics[ width=14cm]{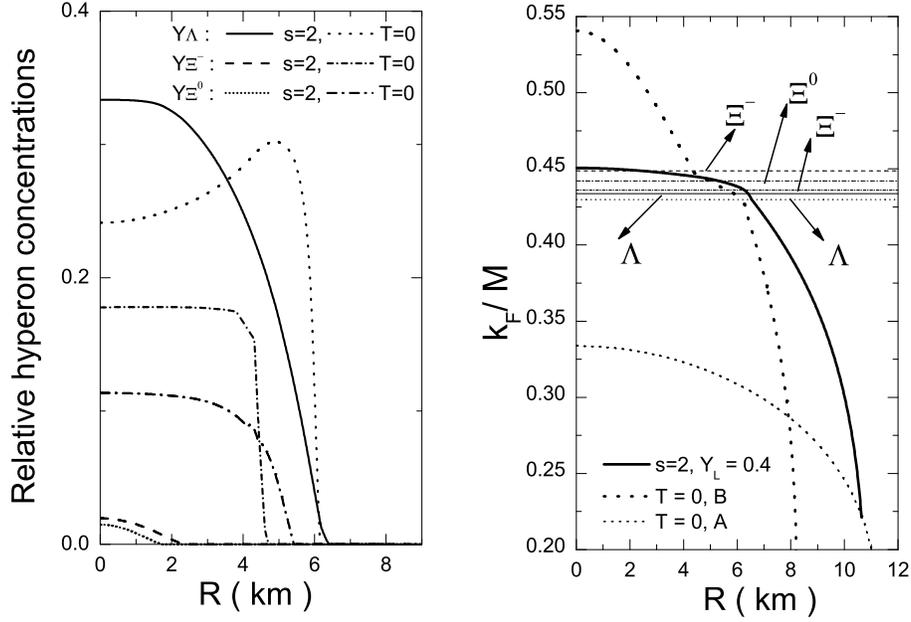}}
 \caption{The fraction of
species $i$, $Y_i$ in the maximum mass configuration as a function
of star radius for the G2 strong parameter set ($k_F$ means the
effective Fermi momentum which specifies the value of $\nu_b$ in
equation (\ref{potchem}). } \label{fig:xaYl}
\end{center}
\end{figure}
\begin{figure}
 \begin{center}
\includegraphics[width=14cm]{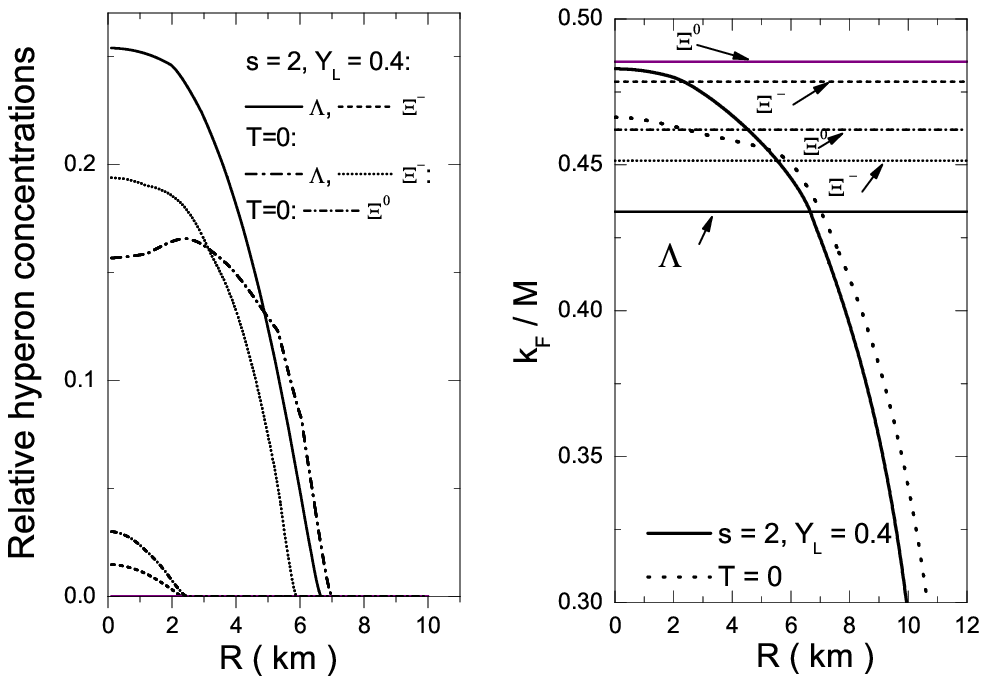}
\caption{The fraction of species $i$, $Y_i$ in the maximum mass
configuration as a function of star radius for the G2 weak
parameter set.} \label{fig:YSdweak}
\end{center}
\end{figure}
 Horizontal lines in the left panels of Fig. \ref{fig:xaYl} and Fig. \ref{fig:YSdweak} correspond to the
threshold density of individual hyperons in the considered model.
The configuration marked as A does not contain hyperons. The
appearance of $\Xi^-$ hyperons through the condition of charge
neutrality affects the lepton fraction and causes a drop in their
contents. Charge neutrality tends to be guaranteed with the
reduced lepton contribution. Results are shown in
Fig.\ref{fig:Ynr}.
\begin{figure}
\subfigure {\includegraphics[width=8.15cm]{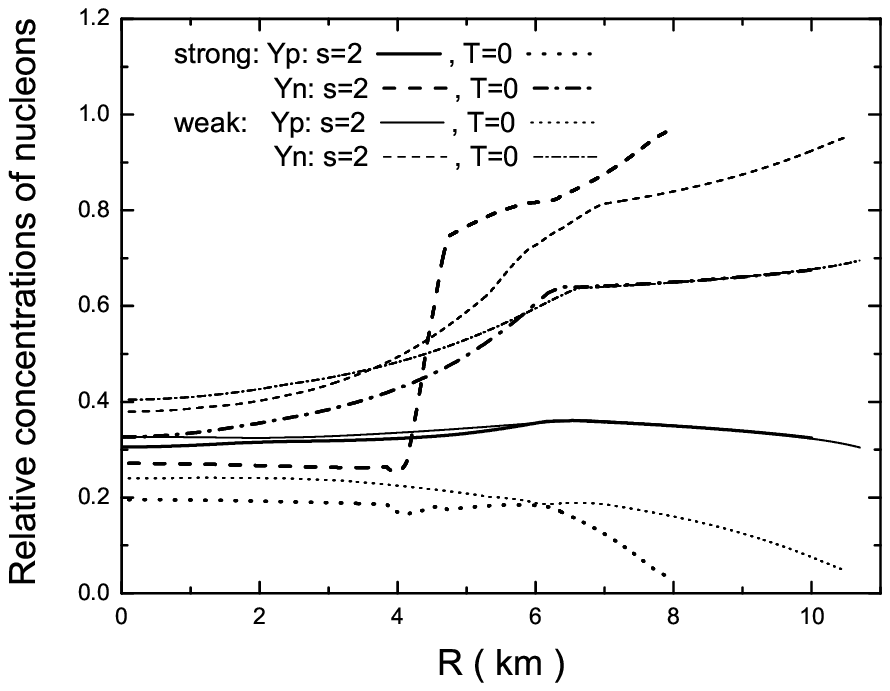}} \subfigure
{\includegraphics[width=8.15cm]{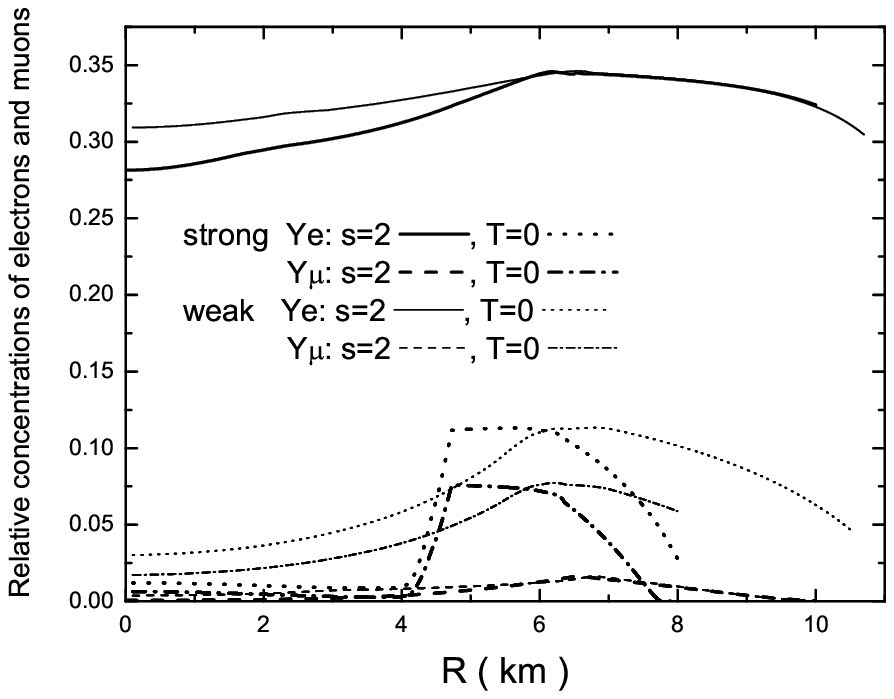}} \caption{Relative
concentrations of $\Lambda$ and $\Xi$ hyperons in  hyperon star
matter for the G2 parameter set.  In the left and right panels the
results for the strong and weak Y-Y interaction, respectively, are
presented.} \label{fig:Ynr}
\end{figure}
  The relative hadron-lepton composition in this model can
be also analyzed through the density dependence of the asymmetry
parameter $fa$ which describes the neutron excess  in the system
and the parameter $fs=(N_{\Lambda}+2N_{\Xi^{-}}+2N_{\Xi^{0}})/n_b$
which is connected with the strangeness contents. Fig.\ref{fig:fa}
and \ref{fig:fs} present both parameters as functions of baryon
number density and the stellar radius $R$. As the density
increases, the asymmetry of the matter decreases and the parameter
$fs$ increases for all the considered cases. Deleptonization and
cooling leads to stellar matter which is more asymmetric and
possesses substantially enhanced strangeness contents.
 \begin{figure}
\subfigure {\includegraphics[width=8.15cm]{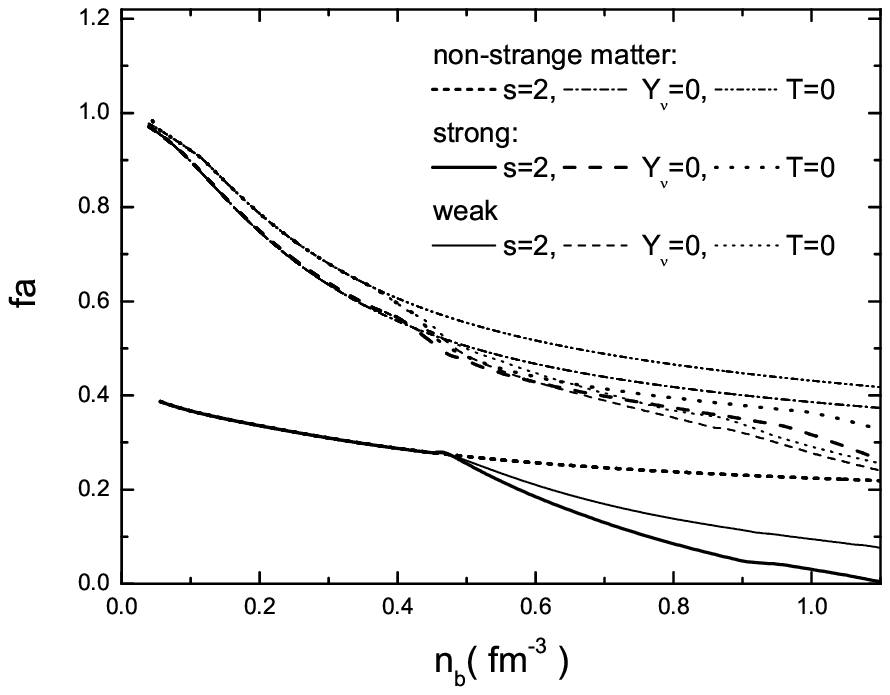}} \subfigure
{\includegraphics[width=8.15cm]{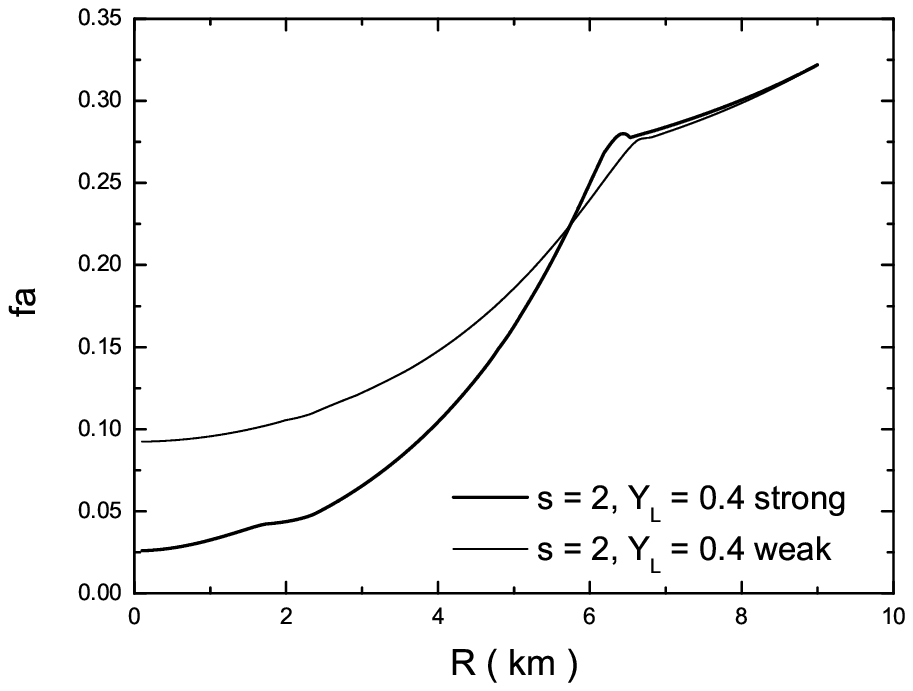}} \caption{The parameter
$fa$ as a function of baryon number density for the strange and
non-strange baryonic matter calculated for each specified
evolutionary phases. The hot models with trapped neutrinos are
marked as $s=2$, whereas hot, deleptonized models as $Y_{\nu}=0$.
In the right panel the parameter $fa$ as a function of stellar
radius for the weak and strong models are presented.}
\label{fig:fa}
\end{figure}
\begin{figure}
\subfigure {\includegraphics[width=8.15cm]{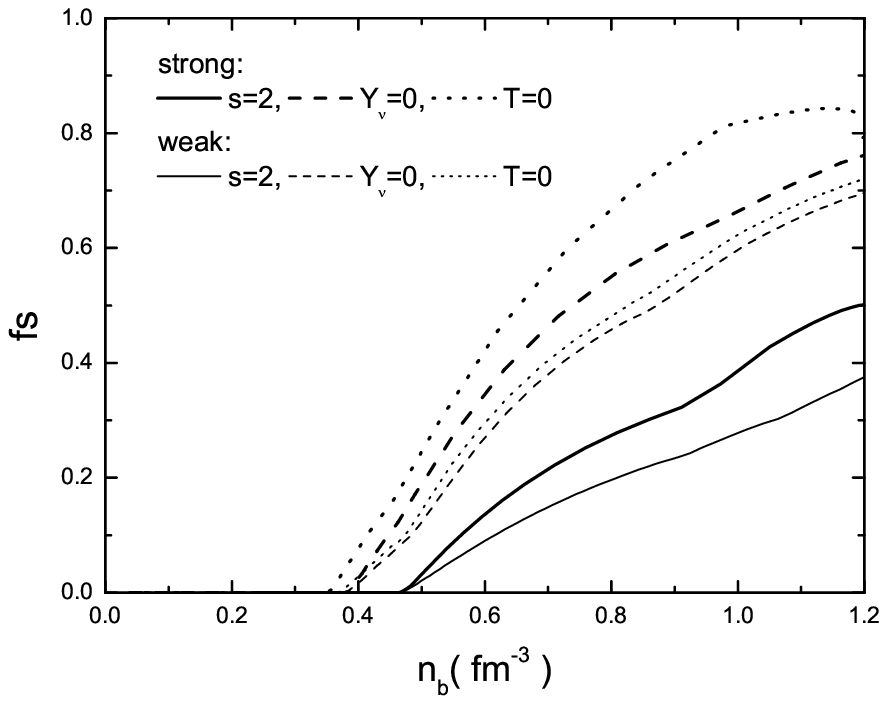}} \subfigure
{\includegraphics[width=8.15cm]{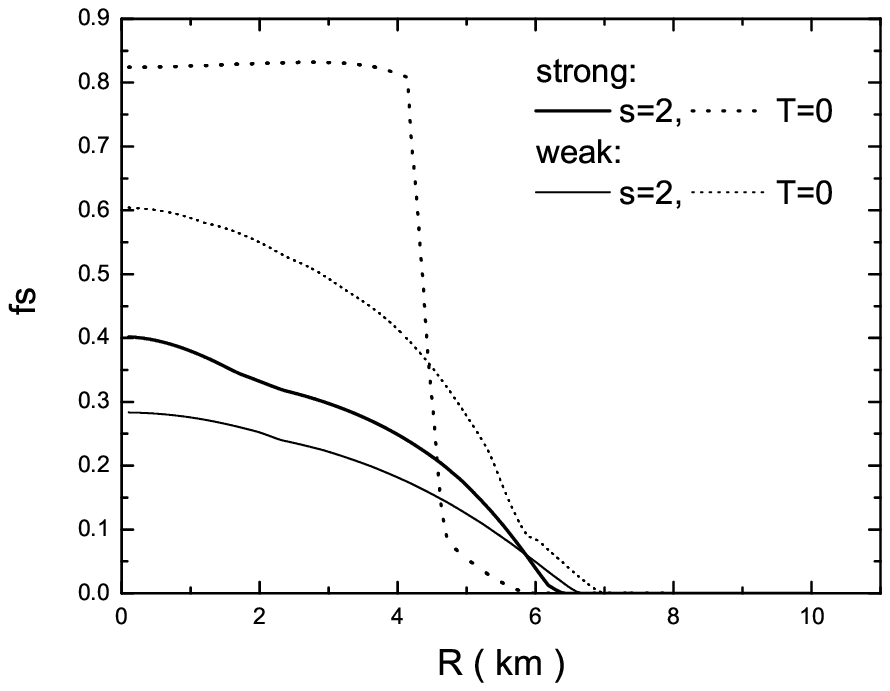}} \caption{The parameter
$fa$ as a function of baryon number density for the strange and
non-strange baryonic matter calculated for each specified
evolutionary phases. The hot models with trapped neutrinos are
marked as $s=2$, whereas hot, deleptonized models as $Y_{\nu}=0$.
In the right panel the parameter $fa$ as a function of stellar
radius for the weak and strong models are presented.}
\label{fig:fs}
\end{figure}
\end{widetext}
\section{Conclusions}
In this paper the complete form of the equation of state of
strangeness rich proto-neutron and neutron star  matter has been
obtained. The considered models are constructed on the assumption
that the proto-neutron star consists of two main parts: the core
and the outer layer. The pressure in the outer envelope is
determined by free nucleons, electrons and neutrinos in $\beta$
equilibrium. The theoretical model of the dense inner core is
described by the Lagrangian which  includes the full octet of
baryons interacting through the exchange of meson fields. The
chosen parameter set based on the effective field theory includes
nonlinear scalar-vector and vector-vector interaction terms which
leads to rather soft equation of state. This is in agrement with
the DBHF results.  The meson sector has been extended not only by
nonlinear terms but also by two additional hidden-strangeness
mesons which reproduce attractive hyperon-hyperon interactions.
Thus the chosen parameter set G2 has been supplemented by the
parameter set related to the strength of the hyperon-nucleon and
hyperon-hyperon interactions. The currently obtained lower value
of the $U_{\Lambda}^{(\Lambda )}$ potential at the level of 5 MeV
permits the existence of additional parameter set which reproduces
this weaker $\Lambda \Lambda$ interaction. The main goal of this
paper was to study the influence of the strength of
hyperon-hyperon interactions on the properties of the
proto-neutron star matter and through this on a proto-neutron star
structure during selected phases of its evolution.  The presence
of hyperons in general leads to the softening of the equation of
state. The behavior of the equation of state is directly connected
with the value of the maximum star mass. Equilibrium conditions,
namely charge neutrality and $\beta$-equilibrium, determine the
composition of the star. It has been shown that replacing the
strong $Y-Y$ interaction model by the weak one introduces large
differences in the composition of a proto-neutron star matter both
in the strange and non-strange sectors. There is a considerable
reduction of $\Lambda$ hyperon concentration whereas the
concentrations of $\Xi^{-}$ and $\Xi^{0}$ hyperons are enhanced
during the deleptonization. In addition, the population of
$\Lambda$ hyperons obtained in the weak models in the considered
evolutionary stages is lower in comparison with those calculated
for the strong $Y-Y$ interaction models.  Thus the weak model with
the reduced $\Lambda$ hyperon population permits larger fractions
of protons and electrons and leads to lower concentrations of
neutrinos. The G2-strong parameterization leads to proto-neutron
neutron stars with the reduced value of the maximum mass. For the
strong hyperon-hyperon interaction strength in the case of the
cold neutron star model besides the ordinary neutron star branch
there exists additional stable branch of solutions which are
characterized by a similar value of the mass but with
significantly reduced radius. This may be connected with the
existence of the third family of stable compact stars \cite{SB}.
The analysis of baryon and lepton concentrations as a function of
stellar radius $R$ shows that for the chosen parameter set there
is no strange baryons in the maximum mass configuration $A$. This
star resembles an ordinary neutron star. The additional stable
configurations of neutron stars which appear on the mass-radius
diagram include large fraction of hyperons (hyperstar). Analysis of the
concentrations of leptons (electrons and muons) in the compact
star $B$ as a stellar radius $R$ leads to the conclusion in the
leptons are gathered? in the outer layer of  the star. Thus, the
star has a lepton rich outer part and almost completely
deleptonized inner core. This analysis shows that there exists a very strong
correlation between the value of the maximum neutron star mass and
the strength of hyperon coupling constants. The inclusion of
additional nonlinear meson interaction terms which modify the high
density behavior of the equation of state together with the strong
hyperon-hyperon interaction lead to the existence of  additional
stable stellar configurations with similar masses and smaller
radii than an ordinary neutron star. The reduction in radius is of
the order of 2.5 km. The internal composition of this additional
neutron star configurations is almost completely free of leptons.
Transmutation similar to the phase transition from the ordinary neutron star
to the more compact hyperstar may be the main origin of the short gamma ray burnst \cite{dar}.\\
 Employing the data concerning the estimated value of the $\Lambda$ well depth
$U_{\Lambda\Lambda}\simeq 5$ MeV the existence of the additional
stable branch of very compact stars have not been confirmed.


\begin{thebibliography}{10}
\bibitem{Shapiro}S. L. Shapiro and S.A. Teukolsky, \textit{Black
Holes, White Dwarfs, and Neutron Stars} (John Wiley , New York,
1983)
\bibitem{Suzuki}M. Fukugita and A. Suzuki \textit{Physics and Astrophysics of
Neutrinos} (Springer-Verlag, Tokyo, 1994)
\bibitem{Prakash}M. Prakash, I. Bombaci, M. Prakash, P. J. Ellis,
J. M. Lattimer and R. Knorren, Phys. Reports \textbf{280}, 1
(1997), \href{http://arxiv.org/abs/nucl-th/9603042}{nucl-th/9603042}
\bibitem{glen}N. K. Glendenning, Astrophys. J.
\textbf{293}, 470 (1985)
\bibitem{glen1}N. K. Glendenning,
\textit{Compact Stars} (Sringer-Verlag, New York, 1997)
\bibitem{georgi1}H. Georgi and A. Manohar, Nucl. Phys.
\textbf{B234}, 189 (1984)
\bibitem{georgi2}H. Georgi, Phys. Lett. \textbf{B298}, 187
(1993)
\bibitem{serot}B. D. Serot, Lect. Note Phys. \textbf{641}, 31 (2004)
\bibitem{bema}I. Bednarek and R. Manka,  Int.Journal Mod.Phys. \textbf{D 10}, 607 (2001)
\bibitem{weber}F. Weber, \textit{Pulsars as Astrophysical Laboratories for Nuclear
and Particle Physics,} (IOP Publishing, Philadelphia, 1999)
\bibitem{gal} J. Schaffner-Bielich and A. Gal, Phys.
Rev. \textbf{C 62}, 034311 (1999)
\bibitem{Schaffner} J. Schaffner, C. B. Dover, A. Gal, C. Greiner, D. J. Millner and H. Stocker, Ann.Phys. \textbf{235},
35 (1994)
\bibitem{FST1}R. J. Furnstahl, B. D. Serot and H. B. Tang, Nucl.
Phys. \textbf{A598},539 (1996)
\bibitem{SerotI} B. D. Serot, Lect.  Notes  Phys. \textbf{641}, 31  (2004),
\href{http://arxiv.org/abs/nucl-th/0405058}{nucl-th/0405058}
\bibitem{FST2}R. J. Furnstahl, B. D. Serot and H. B. Tang,
Nucl. Phys. \textbf{A 615}, 441 (1997),
\href{http://arxiv.org/abs/nucl-th/9611046}{nucl-th/9611046}
\bibitem{sil}T.  Sil, S.  K.  Patra, B.  K.  Sharma, M. Centelles and
X. Vinas, Contribution to \textit{Focus on Quantum Field Theory"}
 (Nova Science Publishers, New York, 2005)
\bibitem{Muller}H.  Muller, Phys. Rev. \textbf{C 59}, 1405 (1999)
\bibitem{Mares}J. Mare$\check{s}$, E. Friedman, A. Gal, and B. K.
Jennings,
  Nucl. Phys. \textbf{A 594}, 311 (1995)
\bibitem{lamsig}C. B. Dover  and A. Gal,  Ann.Phys. \textbf{146},
209 (1983)
\bibitem{ksi}C. J. Batty, E. Friedman, A. Gal and B. K. Jennings,  Phys. Lett.
\textbf{335},
273 (1994)
\bibitem{ksi2}C. B. Dover  and A. Gal, Prog. Theo. Phys. Suppl. \textbf{117},
145 (1994)
\bibitem{Stoks}V. G. J. Stoks and T. S. H. Lee,  Phys. Rev.
\textbf{C 60}, 024006 (1999)
\bibitem{ma}Ma Z.Y., Toki H., B. Q. Chen B. Q. and N. Van Giai, Prog.Theor.
Phys. \textbf{98}, 917 (1997)
\bibitem{shen}H. Shen, Phys.
Rev. \textbf{C 65}, 035802 (2002),
\href{http://arxiv.org/abs/nucl-th/0202030}{nucl-th/0202030}
\bibitem{Takahashi}T. T. Takahashi et
al., Phys.Rev.Lett.  \textbf{87}, 212502 (2001)
\bibitem{bema2} I. Bednarek and R. Manka, J. Phys. G: Nucl. Part.
Phys., \textbf{31}, 1009 (2005), \href{http://arxiv.org/abs/hep-ph/0506059}{hep-ph/0506059}
\bibitem{Song}H. Q. Song, R. K. Su, D. H. Ku and W. L. Qian,
Phys.Rev. \textbf{C 68}, 055201 (2003)
\bibitem{SB}J. Schaffner-Bielich, M. Hanauske, H. St\"{o}cker and
W. Greiner,  Phys.Rev.Lett. \textbf{89} 171101, (2002),
\href{http://arxiv.org/abs/astro-ph/0005490}{astro-ph/0005490}
\bibitem{dar}A. Dar, Hyperstarts - Main Origin of Short Gamma Ray Bursts, \href{http://arxiv.org/abs/astro-ph/0509257}{astro-ph/0509257}
\end{thebibliography}
\end{document}